# Highlights

**Edge effects in radial porosity profiles from μ-CT measurements and melt pool signal intensities for laser powder bed fusion**

Jorrit Voigt, Thomas Bock, Uwe Hilpert, Ralf Hellmann, Michael Moeckel

- Edge enhanced inhomogeneous relative density observed in μ-CT scans
- Multi-layer on axis melt pool monitoring data show effective local process conditions
- Unsupervised time series clustering co-registered with scan pattern shows edge effect
- FEM modeling confirms variations in local thermal histories between edge and center
- Results suggest direction towards tailored modifications of relative density

# Edge effects in radial porosity profiles from μ-CT measurements and melt pool signal intensities for laser powder bed fusion

Jorrit Voigt[a], Thomas Bock[b], Uwe Hilpert[c], Ralf Hellmann[b], Michael Moeckel[a]

[a]Labor für Hybride Modellierung, Technische Hochschule Aschaffenburg, Wuerzburger Straße 45, Aschaffenburg, 63739, Germany
[b]Applied laser and photonics group, Technische Hochschule Aschaffenburg, Wuerzburger Straße 45, Aschaffenburg, 63739, Germany
[c]WENZEL Group GmbH & Co. KG, Werner-Wenzel-Straße, Wiesthal, 97859, Germany

**Abstract**

Limited process control can cause metallurgical defect formation and inhomogeneous relative density in laser powder bed fusion manufactured parts. In this study, cylindrical 15-5 PH stainless steel specimens are investigated by μ-computer tomography; it shows an edge enhanced relative density profile. Additionally, the on axis monitoring signal, obtained from recording the thermal radiation of the melt pool, is considered. Analyzing data for the full duration of the building process results in a statistically increased melt pool signature close to the edge corresponding to the density profile. Edge specific patterns in the on axis signal are found by unsupervised times series clustering. The observations are interpreted using finite element method modeling: For exemplary points at the center and edge it shows different local thermal histories attributed to the chosen laser scan pattern. The results motivate a route towards future design of components with locally dependent material parameters.

*Keywords*: Laser powder bed fusion (L-PBF), μ-computer tomography (μ-CT), melt pool monitoring, machine learning, thermal history, finite element method (FEM)

1. Introduction

Laser Powder Bed Fusion (L-PBF) is an established additive manufacturing (AM) technique in many industrial sectors and can be used for a variety of applications [1–7]. In L-PBF a laser melts metal powder layer by layer resulting in a three-dimensional part. Inhomogeneous defect densities and microstructures within AM manufactured components are a major challenge in L-PBF and limits the application especially in safety-critical components. Pore formation is one common defect and can be categorized into key hole pores and lack-of-fusion defects. Unstable key hole welding modes create spherical pores by gas entrapping [8–10]. Insufficient melting due to lack of energy creates irregular pores [11,12]. Especially these sharp cavities initiate crack growth in cyclic loading [13] and reduces the elongation at break in tensile tests [14]. Further reasons for defect formations are inhomogeneous powder, spatter, and plume induced defects due to insufficient gas flow [12,15]. Several authors reported a change of the layer temperature during the building process, which lead to changes in the relative density and microstructure [16,17]. Usually, the relative density is determined by Archimedes principle or metallographic preparation to evaluate process regimes and the influence of process parameters. A non-destructive method to localize defects in specimens are computer tomography (CT)-scans.

Slotwinski et al. [18] evaluated several density measurement techniques and concluded that a CT scan shows local variability of the porosity in AM manufactured parts. Laquai et al. [19] detected pores by synchrotron X-ray refraction radiography (SXRR) and synchrotron X-ray CT (SXCT) from L-PBF samples to classify them in gas pores and lack-of-fusion defects. The authors concluded, that the SXRR detection and SXCT have similar defect detectability, whereas the SXRR can investigate larger volumes and therefore offers more representative results. Kim et al. [20] introduced a methodology to determine the probability of detection (POD) in X-ray CT scans for AM parts. Bergmann et al. [21] used X-ray CT to measure the relative density of specimens and correlated the findings with optical height measurements. Coeck et al. [22] used CT scans to localize defects in L-PBF manufactured cubes and correlated the positions with anomalies in the melt pool monitoring data.

There have been several studies determine the overall relative density by CT scans, studying the possibility for defect localization, and showing a relationship between melt pool monitoring anomalies and single pore formation, but a detailed analysis of radial porosity profiles and their correlation with an enhanced melt pool signature close to the edge has not yet been published. This study contributes a detailed exploration of the radial relative density profiles for cylindrical specimens. Enhanced edge melt pool signals indicate a change in the process conditions close to the surface, which is correlated with higher relative densities at the edge. A finite element method (FEM) modeling shows different thermal histories between points at the surface and at the center during the process due to the laser scan pattern.

2. Experimental setup

2.1 Laser-based powder bed fusion (L-PBF)

In this study commercially available 15-5 PH stainless steel, which confirms the chemical composition of DIN 1.4540, was used for L-PBF to investigate the local defect formation by μ-CT scans. The used AM machine was an EOS M290 (EOS GmbH, Krailing, Germany) equipped with a 400 W fiber laser (YLR-400-WC, IPG Laser GmbH, Burbach, Germany). For manufacturing a laser power of 170 W, a laser speed of 1085 mm/s, a hatching distance of 0.1 mm, and a layer thickness of 0.02 mm was used. The inert gas was nitrogen ($N_2$), a line scan strategy with a 67 °C rotation without contour hatch was used. The building plate was heated to 100 °C for reducing residual stress.

## 2.2 Computer tomography measurements

To detect and localize pores in the manufactured 15-5 PH stainless steel specimens, an exaCTS90 μ-CT (Wenzel Group GmbH & Co. KG, Wiesthal, Germany) was used. The voltage for the scans was 130 kV with a current of 59 μA. The resolution limit for the metal alloy and experimental setup is a voxel size of 8 μm. Preliminary test with the 15-5 PH stainless steel showed, that a diameter of maximum 3 mm can be investigated with these settings in the μ-CT without beam hardening. The μ-CT samples were manufactured with reference points to align the coordinate systems of the 3D printer and μ-CT specimen (cf. Figure 1 (a)).

For evaluation and transformation of the μ-CT scans VGStudio Max 3.2 (Volume Graphics GmbH, Heidelberg, Germany) software was used and an intensity algorithm (VGDefX) is applied for defect detection. The reference points in Figure 1 (b) are used to assign coordinates to each defect with respect to the position in the samples. In total 10 samples for the lower (14 - 20.5 mm) and higher (44 - 50.5 mm) test heights were investigated. The radial relative density is determined within the rings going from the midpoint to the surface of the cylinder as schematically shown in Figure 1 (c).

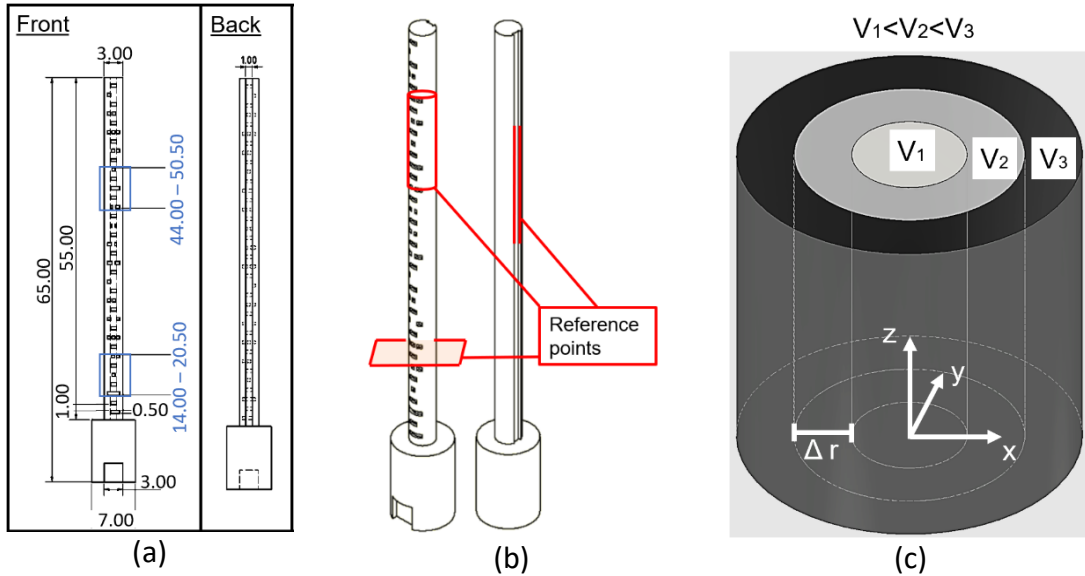

(a) (b) (c)

Figure 1: (a) technical sketch of the manufactured μ-CT samples. (b) schematic illustration of the μ-CT samples with reference points for the x-y-z plane calibration. (c) dissection of the cylinder in rings to determine the radial relative density.

The relative density (RD) for each ring is calculated by equation (1). The step size $\Delta r$ for each ring was 0.2 mm.

$$RD = \frac{V_{ring} - \sum_{i=1}^{N} V_{defect}^{i}}{V_{ring}} \qquad (1)$$

In the equation $V_{ring}$ is the ring volume, $N$ the number of defects within the certain ring, and $V_{defect}^{i}$ the defect volume, respectively.

2.3 Finite element method simulation

In the finite element model, a local heating of a homogeneous and isotropic cylinder is assumed. The laser spot is considered as an internal moving heat source. The solved governing equation is the heat equation:

$$\frac{dT}{dt} = \frac{\lambda}{\rho \cdot C_p}\left(\frac{d^2T}{dx^2} + \frac{d^2T}{dy^2} + \frac{d^2T}{dz^2}\right) \quad (2)$$

where T is the temperature, t represents time, λ the thermal conductivity, ρ is the material density, $C_p$ the specific heat, and Q represents the heat source. The equations for the heat source and the gaussian profile of the laser spot are:

$$\dot{Q} = \frac{2 \cdot P_L}{\pi \cdot r_{spot}^2} \cdot \exp\left(-2 \cdot \frac{r_{focus}^2}{r_{spot}^2}\right) \quad (3)$$

$$r_{focus} = \sqrt{(x - x_{focus})^2 + (y - y_{focus})^2 + (z - z_{focus})^2} \quad (4)$$

The simulated domain and the settings for the heat source (velocity, power, track distance) are the values from the experiments with a laser spot size ($r_{spot}$) of 80 μ m. The phase transitions are modeled by using a smoothed hysteresis of the heat capacity considering the latent heat of fusion at the transition temperatures ($T_{sol}$ and $T_{liq}$). The material properties are shown in Table 1 [23,24]. The simulation environment is COMSOL heat transfer module.

Table 1: Material properties and laser settings for FEM analysis.

| Property | Value |
| --- | --- |
| Density ρ | 7800 kg/m³ |
| Solidus temperature $T_{sol}$ | 1677 K |
| Liquidus temperature $T_{liq}$ | 1713 K |
| Latent heat of fusion $L_f$ | 247 kJ/kg |
| Heat transfer coefficient λ | 13.8 W/m² |
| Heat capacity $C_p$ | 460 J/kg·K |

3. Results and discussion

3.1 CT scan and on axis signal

To study the influence of the height on the porosity, μ-CT scans were made between heights of 14.00 to 20.50 mm and 44.00 to 50.50 mm for 10 specimens each. The relative densities are determined by summing the defect size of the pores in the particular range.

The relative density profiles for the radial distances are determined by equation (1) within the rings of step size $\Delta r = 0.2$ around the center (cf. Figure 1 (c)). The results are presented in Figure 2, in which an increased relative density towards the edge is visible, even though the effect of the radial porosity profiles is more enhanced for the lower test ranges. The decrease of the relative density at 1.4 mm is influenced by measurement inaccuracy in the μ-CT scans due to surface roughness. By distinguishing between the lower and higher test height a moderate trend to an increased relative density at elevated heights is visible.

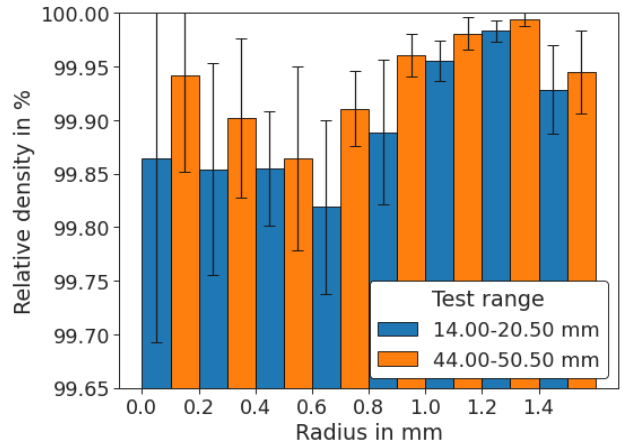

Figure 2: Average relative density with respect to the radial direction in the cylindrical specimens. A trend to larger relative densities at the edge is observed.

The detected defects by the μ-CT investigations indicate a change of the process conditions over radial direction. To monitor this change, the signal of the on axis photodiode is used to analyze the melt pool conditions over height and radius. By comparing the distributions of the on axis signal due to the test height no difference can be observed. In Figure 3 (a), the melt pool signal for a single layer is shown, in which a single, negligible peak can be observed at the edge. However, in Figure 3 (b) the accumulated melt pool data for the full duration of the 20 manufactured samples is shown. At the center the intensity values are lower, whereas a ring of high intensities is clearly visible at the edge. These findings demonstrate a statistical change in the process condition towards the surface, which can not be detected by observing the melt pool signal of a single layer.

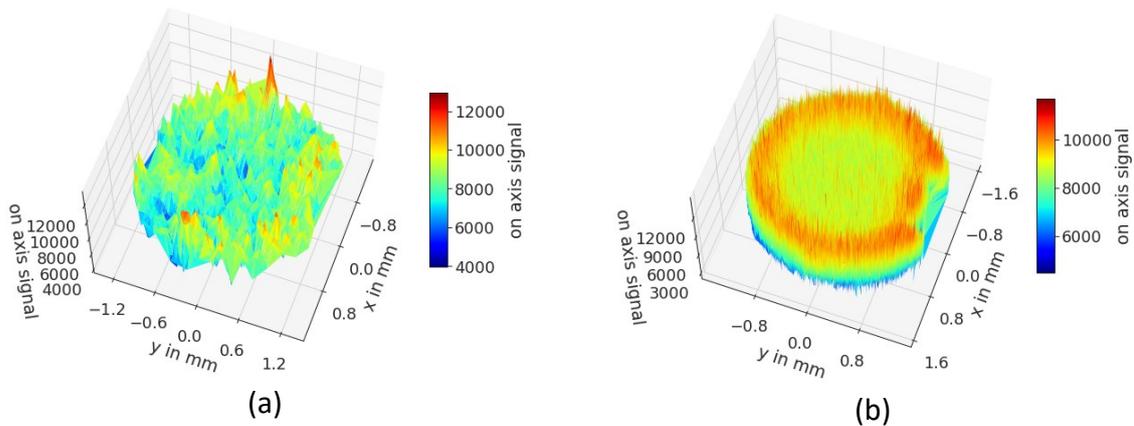

(a)  (b)

Figure 3: (a) the melt pool signals for a single layer with a few peak values can be seen. In (b) the melt pool signal for the full duration of the building process is visible. A significantly enhanced melt pool signature at the edge is observed by investigating the melt pool signal multi-layer-wise.

The trend to a larger relative density at the higher test ranges are in line with observations by Williams et al. [16], who showed a correlation between the dwell temperature at elevated heights and a higher relative density. However, a change in melting conditions can not be detected in the melt pool monitoring data. %The questionable significance in these findings may come from the investigated heights, because the difference in the reported relative density was large for height ranges between 0 to 10 mm and very small for the selected height ranges in this study [16].

The observed radial dependency of the relative density and the signature from melt pool monitoring indicate a smaller likelihood for pore formation close to the surface. It is argued that this is due to locally different effective process conditions since the photodiode signal is linked to the emission intensity of the melt pool. Hence it has been argued that it shows a correlation with the melt pool temperature [25]. Variations in the photodiode on axis signal indicate locally different thermal conditions caused for instance by heat accumulation. To classify the edge melt pool signals, the DBSCAN clustering algorithm implemented with the python environment scikit-learn is used. Time series sections defined by 10 measurement points around anomalies of the on axis photodiode signal ($\leq$ 12000 units) are normalized and inspected. Naive black box clustering results in a set of various apparently similar classes and a residual class of noisy data. After calibration (eps=0.55, $\min_{samples}$=5), three classes with different signatures are considered: a first class is characterized by a sudden increase and fast subsequent relaxation of the photodiode signal (Fig. Figure 4 (a)). Co-registration with the spatial trajectory of the laser path shows that these events occur at turning points after re-entrance of the laser focus. This can be explained by preheating of the site due to the nearby recent passage of the laser focus. FEM simulations for thermal histories of four exemplary points confirm this effect (see section 3.2). This effect can be observed at lower on axis signal levels, too. However, noisy data makes a statistical significant observation more difficult.

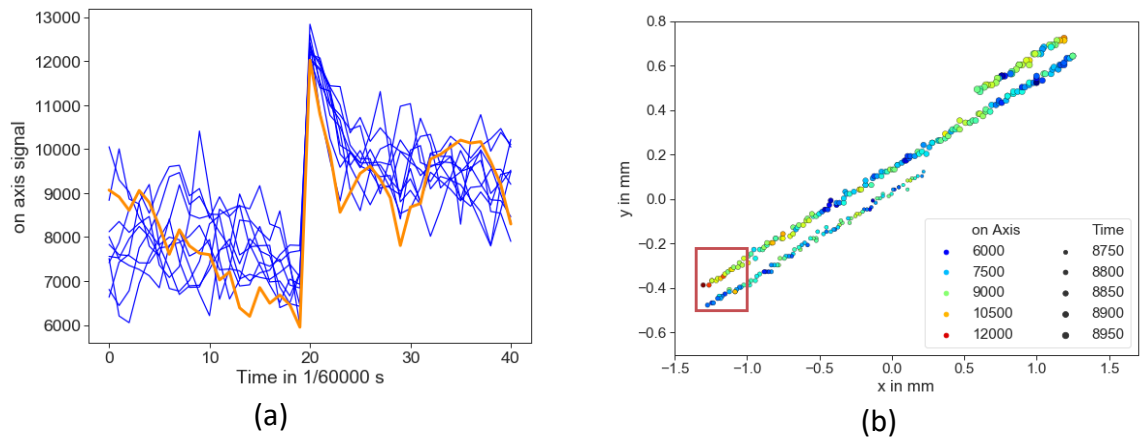

(a) (b)

Figure 4: (a) examples for class 1 of the time series clustering are characterized by a sudden increase and subsequent relaxation of the melt pool signal. In (b) the location of one such event (orange line in (a)) on the laser trajectory is shown. The anomaly clearly resides at the re-entry of the laser focus after a turning.

A second class describes similar patterns in a neighborhood of the boundary with a smoother increase in the on axis signal. A residual class contains all other noisy time series data.

3.2 Finite element method for thermal histories

To simulate the temperature at four exemplary points at the center (Point 1/4) and at the edge (Point 2/3) of the cylindrical specimen, the finite element method is used. At each point a direct and a subsequent nearby passage of the laser focus are considered, see Figure 5 (a). The multiphysics simulation includes the heat conduction in the solid material and powder as well as the phase changes. The gaussian laser focus has a spot size of 80 µm and linear trajectories are assumed with a hatch distance of 100 µm.

The results are shown in Figure 5 (b). For the points at the center (Point 1/4) the temperature peaks after the first passage of the laser focus and decays below the liquidus temperature, while a second peak occurs clearly separated from the earlier one. For the points at the edge (Point 2/3) a superposition of two temperature peaks occur due to the shorter time difference between the passages. When the laser focus approaches the edge (e. g. Point 2), a first peak caused by direct passage of the laser focus is followed by a lower post-heating peak.

Post-heating occurs due to a second nearby passage of the laser focus. The superposition of both passages leads to an extended temperature history above liquidus. However, when the laser focus departs from the edge (e. g. Point 3), the temperature profile shows a first lower pre-heating peak due to the nearby passage of the laser focus and a subsequent larger peak. As a consequence of the pre-heating this second peak surpasses the maximum temperatures from all other points in this analysis. This elevated temperature gives rise to enhanced thermal signatures.

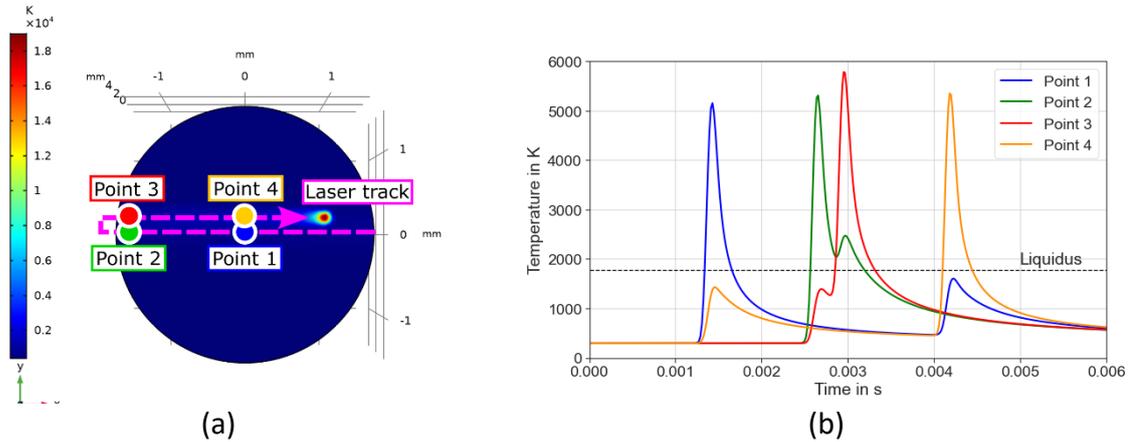

Figure 5: (a) illustrates the top view of the domain and the laser scan pattern. Four points of interest are at the centre of the sample (Point 1/4) and at the edge (Point 2/3). In (b) the thermal histories for the points are shown. Two clearly separated temperature peaks are visible for Point 1/4, while both peaks are merged for Point 2/3. At Point 2 liquidus is surpassed for a prolonged time. At Point 3 pre-heating leads to an enhanced maximum temperature.

The extended thermal history for Point 2 exceeds the liquidus on a prolonged scale compared to the thermal history of Point 1/4. Furthermore, the time integral of temperature above liquidus for Point 2 surpasses that one of Point 1/4. In consequence, the melting process is enhanced and gives rise to less gas entrapping and reduced concentration of lack of fusion defects at the edge, which is a possible explanation for the changing radial defect densities (cf. Figure 2).

The higher melt pool intensity (cf. Figure 3) for Point 3 can be explained by a pre-heating effect caused by previous laser passages through neighboring locations. The emitted intensity (I) of a body is correlated with the temperature (T) according to the Stefan-Boltzmann equation ($I \propto T^4$), which leads to significant changes in the signal even at comparatively small temperature changes.

4. Conclusion

In this study, the local fluctuations of the relative density for L-PBF manufactured cylindrical specimens are explored by µ-CT scans. The relative density for the cylindrical specimens showed an increase towards the edge and a slight trend for elevated heights. Furthermore, melt pool monitoring has been conducted based on the photodiode on axis signal. Increased intensities of the melt pool signal is observed primarily at the edge. Time series sequences related to peak intensities are classified into different categories by unsupervised machine learning techniques. The results indicate varying effective process conditions close to the surface.

FEM simulation is conducted to explain the observed behavior by depicting the local thermal histories for different exemplary points at the center and edge for a given laser scan pattern. The analysis demonstrates that at points close to the edge modified thermal profiles due to peak superposition appear: Either (1) prolonged relaxation time above liquidus caused by post-heating from a subsequent laser passage or (2) elevated peak temperatures as a consequence of pre-heating by the previous laser passage.

In conclusion, elevated peak temperatures caused by a given scan pattern give rise to thermally enhanced melt pool signatures and to a presumed reduction of lack of fusion defects, visible in μ-CT measurements as increased relative density close to the surface. The results suggest further work on thermal control based on scan pattern definitions and corresponding *in-situ* monitoring to optimize and design locally varying material parameters, e. g. the relative density.

## 5. Acknowledgement

The Laboratory for Hybrid Modeling receives funding from the Bavarian State Ministry for Science and Arts for the project PromoAdd3D. Author Contributions: Data analysis, Modeling, Writing, Visualization-J.V.; Experimental measurements (μ-CT), Visualization-T.B.; μ-CT analysis-W.H.; Supervision, project administration, funding acquisition, review-R.H. and M.M.